# Who Gets Recommended? Investigating Gender, Race, and Country Disparities in Paper Recommendations from Large Language Models

Yifan Tian

*Department of Information Management, Peking University, Beijing 100871, China*

Yixin Liu

*Department of Information Management, Peking University, Beijing 100871, China*

Yi Bu

*Department of Information Management, Peking University, Beijing 100871, China*
*Center for Informationalization and Information Management Research, Peking University, Beijing 100871, China*
*Center for Digital Intelligence Science and Education Research, Peking University Chongqing Research Institute of Big Data, Chongqing 401332, China*

Jiqun Liu

*School of Library and Information Studies, The University of Oklahoma, Norman, Oklahoma 73019, U.S.A.*

**Note:** Yifan Tian and Yixin Liu contributed to this paper equally.

**Correspondence:** Corresponding concerning this article should be addressed to Yi Bu (buyi@pku.edu.cn) and Jiqun Liu (jiqunliu@ou.edu).

**Abstract**: This paper investigates the performance of several representative large models in the tasks of literature recommendation and explores potential biases in research exposure. The results indicate that not only LLMs' overall recommendation accuracy remains limited but also the models tend to recommend literature with greater citation counts, later publication date, and larger author teams. Yet, in scholar recommendation tasks, there is no evidence that LLMs disproportionately recommend male, white, or developed-country authors, contrasting with patterns of known human biases. Our study explores the potential biases in LLM's literature recommendation, a growing need and application in scholarly research and communication, employs quantitative methods to uncover the limitations and potential biases of LLMs in literature recommendation, offering valuable insights for the development of fairer and more effective academic recommendation tools in the future.

## INTRODUCTION

In the field of Science of Science, research on large language models (LLMs) has focused on applications such as academic text mining, writing assistance, research evaluation[1–3], automated literature reviews[4], data annotation[5], reference generation[6], and peer review[7], which significantly reduces the manual workload for researchers. By streamlining these tasks, LLMs enable more efficient handling of scientific information and broaden access to academic resources, potentially lowering barriers for researchers worldwide, especially for individuals from non-English-speaking regions and populations. However, the integration of LLMs into scholarly work raises ethical concerns[8], particularly around authorship, intellectual property, and the risk of reinforcing biases present in academic content[9,10]. As LLMs continue to shape scientific practices, it is crucial to investigate their impact on research norms, integrity, and equity within the academic community.

While the promising applications of Large Language Models (LLMs) is widely acknowledged, the ethical issues associated with their deployment in academic research and communication have sparked ongoing and vigorous debate. Key concerns include the attribution of authorship in academic writing, where the distinction between human and machine contributions becomes increasingly ambiguous, raising complex questions regarding intellectual property rights and inequality in access and exposure. This ambiguity has led some researchers to firmly oppose listing LLMs, such as ChatGPT, as co-authors on scholarly publications[11,12]. Furthermore, the integration of LLMs into educational settings could transform teaching methodologies and reshape students' learning, interest development, and information interaction experiences, potentially challenging traditional pedagogical approaches. Academic integrity is also at stake, as LLMs may unintentionally perpetuate biases embedded within their training data, thus introducing risks of academic discrimination. In the academic domain, these ethical concerns are particularly pressing, highlighting the need for a thoughtful and balanced approach to LLM utilization that respects the core principles of academic rigor and fairness.

Therefore, particularly in the social sciences, bias and discrimination in LLMs has emerged as a critical research topic. Gallegos et al.[13] have explored this issue extensively, defined social biases in natural language processing, developed metrics to evaluate these biases, and proposed mitigation techniques to address biases related to gender and social groups. This area of study attracts attention from the scientometrics community as well. For instance, Petiska's study[12] tasked ChatGPT with writing literature reviews in ten environmental science subfields and analyzing the characteristics of the cited literature. The study found that ChatGPT tends to favor highly cited, older publications from renowned journals, suggesting a potential amplification of the Matthew Effect or existing biases in popularity and research exposure within the discipline.

These preliminary findings naturally lead to further questions:

**RQ1**: To what extent do LLMs exacerbate the Matthew Effect in other disciplines?

**RQ2**: To what extent do LLMs intensify other dimensions of inequality within the scientific community?

Exploring these questions further could involve using LLMs in literature recommendation tasks to uncover any existing biases and determine the extent of these biases. In the field of science of science, many existing works have addressed the disparity in research exposure. The key factors can be majorly categorized into four main areas, namely citation counts, gender, race, and country. A significant number of studies report the existence of the Matthew effect in the citation and dissemination of research outcomes[14,15]. In other words, people are more likely to disseminate or cite articles that have already been cited or disseminated more frequently. Larivière et al.[16] and many other studies[17,18] have reported that the works of women face systemic bias in citation, dissemination, and recognition. Reports by Hopkins et al.[19] have highlighted that the research of scholars from racial minorities are unfairly treated in terms of citation and media mention. Additionally, compared to scholars from developed countries conducting similar research, scholars from developing countries receive fewer citations and less media coverage[20] and occupy supporting rather than leading roles[21]. By analyzing recommendation tendencies, researchers can better measure the potential biases of LLMs and understand their impact on academic equity and fairness in scholarly communication, research exposure and representation. Findings from this study can also inform the design of policies and regulations on LLM applications in academic search and recommendation.

## DATA

In this paper, we primarily utilize two datasets: one containing the key papers recommended by LLMs within specific fields, including Machine Learning, Reinforcement Learning, Deep Learning, and Natural Language Processing, and the other comprising the "real" important papers within these fields. The former dataset was generated through prompt engineering using the APIs provided by various LLMs, while the latter consists of papers manually curated based on citation counts and domain expertise (See Appendix). The bibliographic information for the papers and authors used in this study is sourced mainly from SciSciNet[22] and OpenAlex[23]. Detailed descriptions of the prompts used and the methods for calculating the various metrics can be found in the METHODS section.

## METHODS

The flowchart in Figure 1 outlines the methodology for the experimental process described in the paper. The study begins by determining the focus field, i.e., Machine Learning (ML), Deep Learning (DL), Reinforcement Learning (RL), or Natural Language Processing

(NLP). Next, a carefully crafted prompt is designed to guide the language models in generating outputs relevant to the selected field. This prompt is tested with multiple LLMs in multiple rounds, including GPT, GLM, and Claude, to generate lists of important papers or authors that the models recommend as significant to the field. These recommendations represent the machine-generated perspective. In parallel, a list of 'real' important papers or authors is compiled based on domain expert knowledge. This step serves as a benchmark for assessing the accuracy and alignment of the LLM-generated recommendations. The two lists—one generated by the LLMs and the other determined by experts—are then compared through hypothesis testing. This involves analyzing differences in various attributes between the two lists to identify patterns, discrepancies, or potential misalignments. Finally, the research hypotheses are validated, and any new potential biases in the recommendations of LLMs are identified. By repeating the experiment and refining the analysis, the study aims to provide insights into the reliability and biases of LLMs in identifying significant literature within a given research domain. This approach contributes to a deeper understanding of the strengths and limitations of LLMs in scientific knowledge discovery.

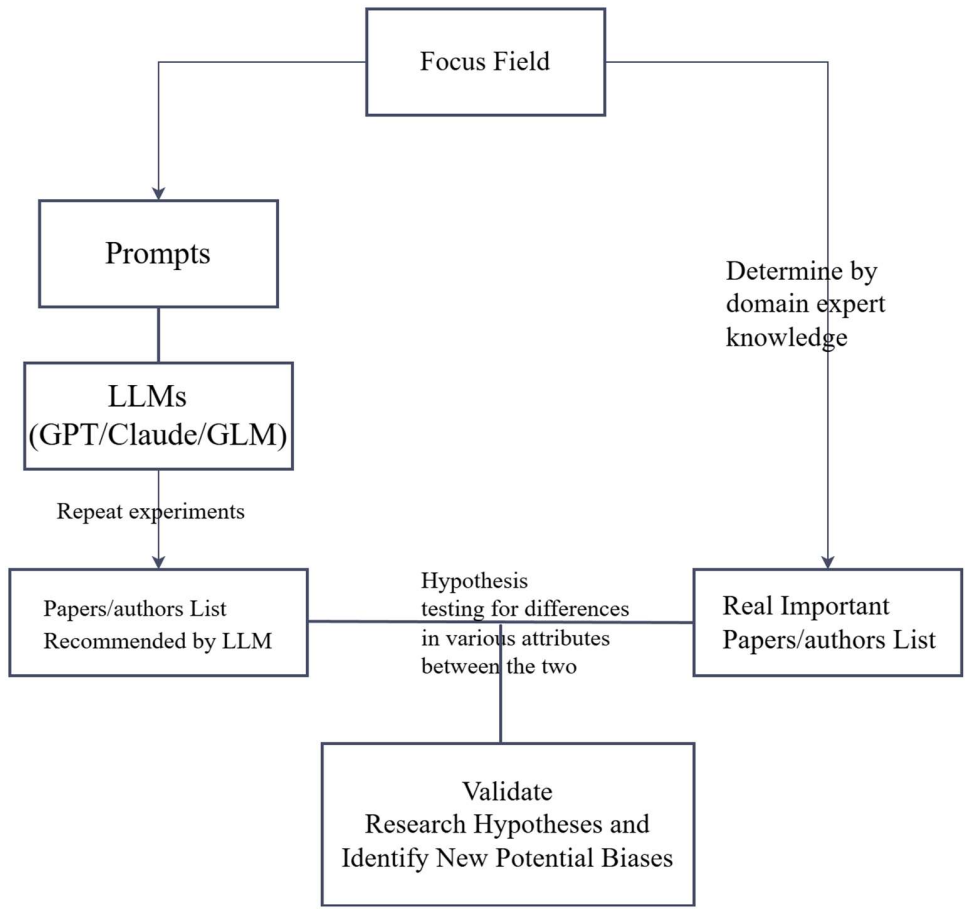

**Figure 1. Flowchart of the methodology for the experimental process**

*Models*

For the empirical analysis, we selected three prominent LLMs: ChatGPT (*https://cdn.openai.com/papers/gpt-4.pdf*), Claude (*https://www.anthropic.com/constitutional-ai*), and GLM[24]. ChatGPT, developed by OpenAI, is a highly advanced language model recognized for its exceptional performance and widespread use across various natural language processing tasks[25,26]. Claude, introduced by Anthropic, prioritizes safety and reliability, excelling in text generation and comprehension. GLM, developed by Tsinghua University in China, is tailored for Chinese corpora and exhibits strong adaptability. These models, originating from leading technology companies and institutions, embody diverse design philosophies, training datasets, and application scenarios, offering a comprehensive representation of the performance and potential biases of widely-applied, potentially high-impact LLMs in the literature recommendation task. To minimize the influence of randomness in the models' responses, multiple queries were conducted for each task.

*Task and Prompt*

This study aims to explore the performance of LLMs in literature recommendation and their potential biases. The specific inquiry methods are in twofold: (1) Important paper recommendation: Inquire the LLM to assume itself to be an expert in the field and recommend 50 important/important papers in the field. (2) Important scholar recommendation: Inquire the LLM, assuming it to be an expert in a specific field, to recommend 50 important scholars in the field (also requiring the LLM to provide a corresponding work for each scholar to avoid hallucination and facilitate alignment).

Additionally, 50 actual important papers and 50 actual important scholars in the field were identified through citation sorting and manual screening. We initially selected the fields of artificial intelligence and machine learning, exploring a variety of topics with different scopes and levels to assess LLMs' accuracy in reflecting real-world situations. These topics included fundamental concepts such as machine learning and deep learning, as well as interdisciplinary concepts like AI in healthcare and ethical AI. However, after preliminary explorations, we found that most LLMs performed poorly on interdisciplinary concepts, as most of the essays generated were proven to be fake. Consequently, after evaluating the performance on these topics, we ultimately decided to focus on four fundamental concepts: machine learning, deep learning, reinforcement learning, and natural language processing, as our primary research fields. These topics demonstrated better performance in providing accurate recommendations and encompassing different levels of detail; for example, machine learning offers a broader perspective while deep learning focuses on more specific aspects. API inquiries were conducted separately for each subfield with the following prompts.

Important paper recommendation prompt content:

*"You are an AI who is knowledgeable in academic fields as well as AI and willing to help people. I would like to understand how research in the field of {machine learning} is like. Can you recommend to me 50 real essays in the field of machine learning? The amount shouldn't be more than 50 or less than 50 and all of them must be real."*

Important scholar recommendation prompt content:

*"You are an AI who is knowledgeable in academic fields as well as AI and willing to help people. I would like to understand how research in the field of {machine learning} is like. Can you recommend to me 50 real scholars in the field of machine learning with one of their most famous essays? The amount shouldn't be more than 50 or less than 50 and all of them must be real."*

Previous research on LLMs has shown that their responses exhibit a certain degree of randomness[27]. Additionally, some studies have reported that interacting with LLMs at different time periods can lead to varying results[28]. To eliminate the influence of inherent randomness and interaction timing, we repeated each experiment with the LLMs three times at different time points. Specifically, we selected three time slots (Beijing Time, GMT+8): 4 pm, 9 pm, and 11 am, to cover the primary working hours across as many regions as possible. The results from these multiple experiments were aggregated, and the average outcomes are presented in the subsequent sections.

*Research Article Retrieval*

We retrieved citation counts, author details, topic classifications, as well as scholar citation counts, institutional affiliations, and country information using the OpenAlex API. OpenAlex is a free and open platform designed to provide seamless access to academic information, building upon and extending the capabilities of the Microsoft Academic Graph (MAG)[29]. It encompasses a vast array of academic data, including publications, authors, institutions, journals, and scholarly concepts, offering researchers a comprehensive and high-quality resource for academic studies. Due to its rich dataset and versatility, OpenAlex has been widely adopted to address a variety of research questions[30–32].

*Interdisciplinarity and Disruptiveness Calculation*

Interdisciplinarity is a multifaceted concept with various strategies for measurement. In bibliometrics, it is commonly examined through the lens of diversity, which encompasses three key dimensions: variety, balance, and disparity[33]. Variety refers to the number of disciplines represented in a publication's references; publications citing a broader range of disciplines typically exhibit higher variety. Balance assesses the distribution of references across these disciplines; a highly uneven distribution results in lower balance, indicating less uniform interdisciplinarity. Disparity measures the semantic differences between disciplines cited in the references; references from disciplines that are semantically distant contribute to greater disparity. Together, these dimensions provide a nuanced understanding of interdisciplinarity in scholarly work.

The domain of bibliometrics and scientometrics has already proposed many indicators and variations to operationalize the above-mentioned three dimensions for understanding interdisciplinarity. Among these, DIV is a commonly adopted measure that incorporate the three dimensions into a whole[34]. Specifically, we use the "field" attribute in the OpenAlex dataset to represent the domain of each paper. OpenAlex categorizes all papers in the dataset into 26 fields. With this classification and the citation relationships between papers, we can easily calculate the DIV index of the papers:

$$DIV = variety \times balance \times disparity \tag{1}$$

$$Variety = \frac{n}{N} \tag{2}$$

where $n$ is the number of field categories in the references of the key literature, and $N$ is the total number of fields in the primary classification system.

$$balance = 1 - Gini = 1 - \frac{\sum_i^n (2i - n - 1)x_i}{n \sum_i^n x_i} \tag{3}$$

where $n$ is the number of domain categories, $i$ is the index of the domain categories sorted in non-decreasing (increasing) order, and $x_i$ is the number of references in the $i$-th domain category.

$$disparity = \sum_{ij(i \neq j)}^n \frac{d_{ij}}{[n \times (n-1)]} \tag{4}$$

Where $n$ is the number of domain categories included in the references of key papers, and $d_{ij} = 1 - cos(row_i, row_j)$ represents the difference between domain $i$ and domain $j$ (1 minus the cosine similarity between any two rows in the citation matrix).

The disruptiveness index was obtained from the SciSciNet dataset, an open academic database designed for scientometric research, supporting the construction and analysis of scientific knowledge maps, and containing commonly used scientometric indicators, including the disruptiveness index defined by Wu et al.[35].

*National, Racial, and Gender Information Retrieval*

The World Bank's Human Development Index (HDI) data was used as an indicator of national development levels. Countries with "Very High HDI" were defined as developed countries while the others were categorized into developing countries.

*Evaluating Recommendation Results*

To verify the authenticity of the recommended articles and authors, we manually searched the recommended titles in OpenAlex. Recommendations were deemed false if the edit distance between the recommended title and the top result in OpenAlex exceeded a threshold determined through experience. Additionally, articles with fewer than 100 citations were also considered false, given the relatively high average citation counts in the field of computer science.

The distributions of various characteristics of the recommended literature and scholars from each LLM were then compared against the actual distributions to identify discrepancies. Kolmogorov-Smirnov tests and Chi-Square tests were performed to assess statistical differences between the recommendations and the actual distributions, supplemented by Kernel Density Estimate (KDE) plots for visual analysis.

*Hypotheses*

To clarify the core focus of our study (RQ1 and RQ2), we propose four hypotheses based on the biases currently observed in the scientific community. Our experiments are designed to test these four key hypotheses one by one. It is important to note that, beyond testing these hypotheses, this study also provides original insights in other areas, such as the age of the literature, team size, and more.

In the science of science, many studies have examined disparities in research exposure, focusing on four main factors: citation counts, gender, race, and country. The Matthew effect highlights a tendency to favor already widely cited works[14,15] (Hypothesis 1). Systemic biases have been reported against women in citations and recognition[16–18] (Hypothesis 2) and racial minorities in citations and media mentions[19] (Hypothesis 3). Scholars from developing countries also face reduced citations and coverage compared to those in developed countries[20,21] (Hypothesis 4). By testing these four key hypotheses, we explore whether LLMs exhibit potential biases in literature recommendation tasks.

**Table 1. The four key hypotheses.**

|   | Hypothesis |
|---|---|
| 1 | LLM tends to recommend papers with higher citation counts. |
| 2 | LLM tends to recommend male scholars. |
| 3 | LLM tends to recommend white scholars. |
| 4 | LLM tends to recommend scholars from developed countries. |

# RESULTS

*Error rate*

We began by evaluating the authenticity of the recommended papers to determine whether they were real. The error rates of the LLM recommendations varied, as shown in Table 1, which presents the recall and precision metrics for each model. Among the three, Claude demonstrated the best performance, while GLM showed the poorest. Due to GLM's high error rate and the fact that most of its recommended results were low-citation documents with limited reference value, as observed in Table 1, we excluded GLM's outputs from subsequent visualizations. This decision reflects the model's inadequate performance in providing reliable recommendations.

**Table 2. The recall and precision of the recommendations results by each LLM.**

|   | Precision (%) | Recall (%) |
|---|---|---|
| ChatGPT-4o | 15.20 | 16.58 |
| Claude | 20.00 | 24.06 |
| GLM | 3.19 | 4.28 |

*Paper Level Bias*

The average citation count of documents recommended by LLMs is lower than that of actual important documents. However, the similar distribution in the high-citation range suggests that LLMs do consider highly cited documents, albeit with less emphasis on citation count as a primary indicator of quality compared to real-world benchmarks. Regarding team size, the documents recommended by LLMs closely align with real patterns, with both primarily comprising small teams of fewer than 10 members. In terms of document age, LLMs show a significant preference for recently published works, particularly those from the last two decades.

When examining disruptiveness, while most of both LLM recommendations and real documents predominantly feature works in Machine Learning field with a disruption score below 0.5, the average disruptiveness of real important documents is higher. This indicates that LLM recommendations may tend to be more conservative, favoring developmental papers that build on existing research directions rather than highly disruptive works. For interdisciplinarity, LLM recommendations generally resemble the actual situation, with only 2 cases where the interdisciplinarity of LLM recommended papers are significantly lower. This alignment might reflect the tendency for narrower subfield spans to minimize communication costs, thereby enhancing the quality of academic outputs.

In the reinforcement learning field (Figure 1B), LLMs continue to show a preference for newer documents. However, there are no distinct preferences regarding team size, disruptiveness, or interdisciplinarity in this domain. LLMs do not exhibit a strong preference for highly cited documents in their recommendations, covering a wide range of citation counts from low to high. They prioritize recent publications and those focused on developmental research. This tendency could stem from the relatively recent training data of LLMs and their aim to adhere closely to user-specified fields when generating recommendations.

In the deep learning field (Figure 1C), apart from differences in citation count, the variations between LLM recommendations and real results are minimal. This may be attributed to the relatively short history of deep learning and its rapid development, where a few foundational documents have established widely recognized structures.

In the natural language processing (NLP) field (Figure 1D), LLMs exhibit a preference for documents produced by larger teams. As in other fields, they also favor more recent publications and conservative, developmental works. In terms of interdisciplinarity, LLM recommendations show no notable deviation from the real distribution.

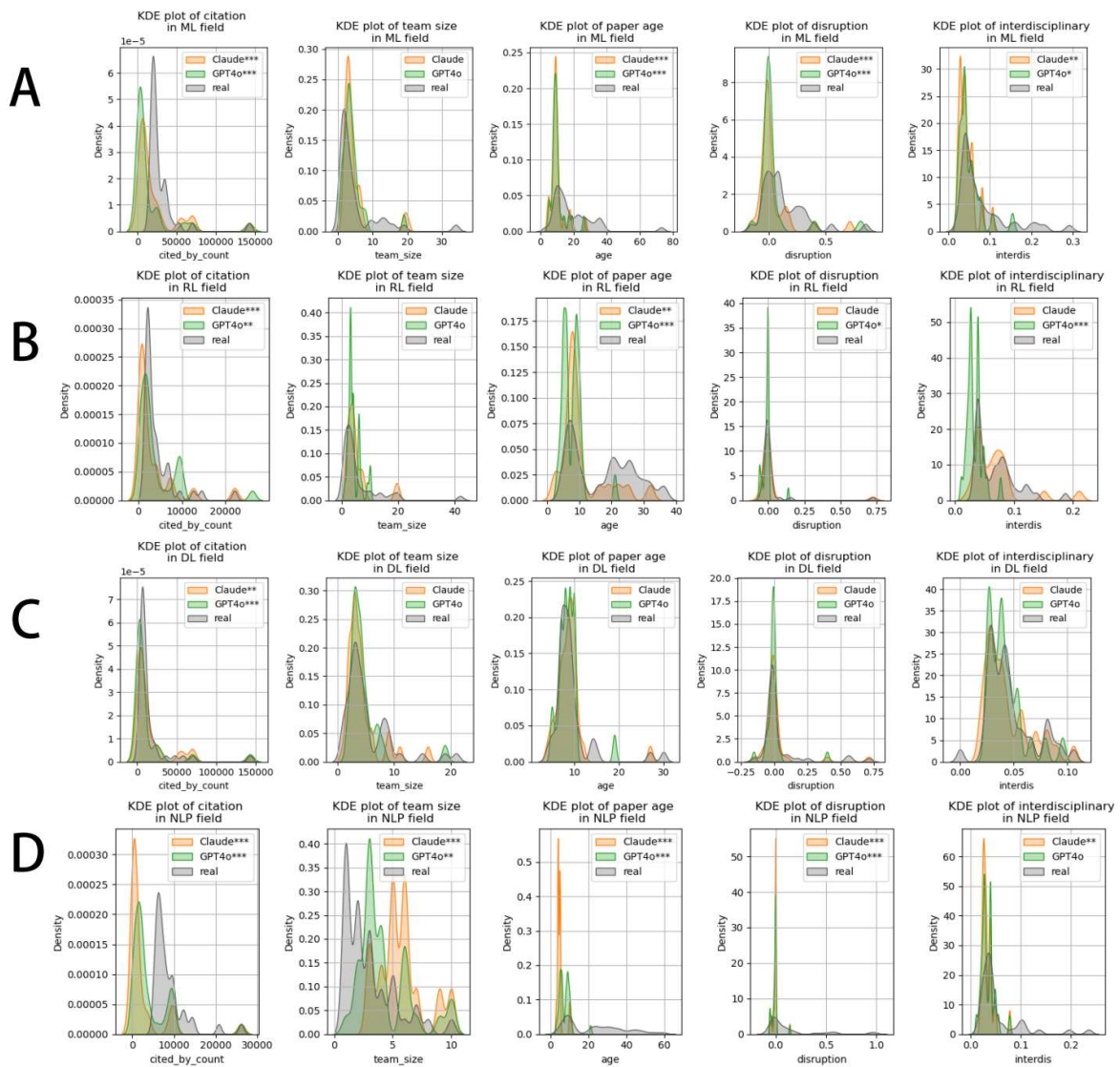

Figure 2. Kernel Density Estimation (KDE) of LLM recommendation results versus real situation across citation counts, team size, paper age, disruption and variety of recommended papers in the 4 main field. The asterisks in the legend represent the significance of the difference compared to the actual situation (judged by the Kolmogorov-Smirnov test, ***: $p<0.01$, **: $p<0.05$, *: $p<0.1$, no asterisk indicates no significant difference from the actual situation). Since these are KDE results, the graphs for citation, team size, etc., may show images on the left side of 0.

We further examined the differences between the topic distribution of recommendations made by large language models and the actual distribution of research topics in the real world. Specifically, we used the subfield classifications of articles from the OpenAlex dataset as a representation of research topics across four major domains: Machine Learning (ML), Deep Learning (DL), Natural Language Processing (NLP), and Reinforcement Learning (RL). A weighted network was constructed based on the co-occurrence of subfields, and an embedding model was employed to calculate the similarity between subfields using their descriptive metadata. These similarity scores were then used to adjust the edge weights in the network.

To visualize the differences, overlay graphs were generated for the four domains, comparing the topic distributions in LLM recommendations (ChatGPT-4, Claude) with the actual distributions in the literature (Figures 2, 3, 4, and 5). Overall, the topic coverage of LLM recommendations was found to be partial, capturing only a subset of the topic landscape within each domain. Additionally, subtle differences in topic emphasis were observed between the recommendations provided by different LLMs.

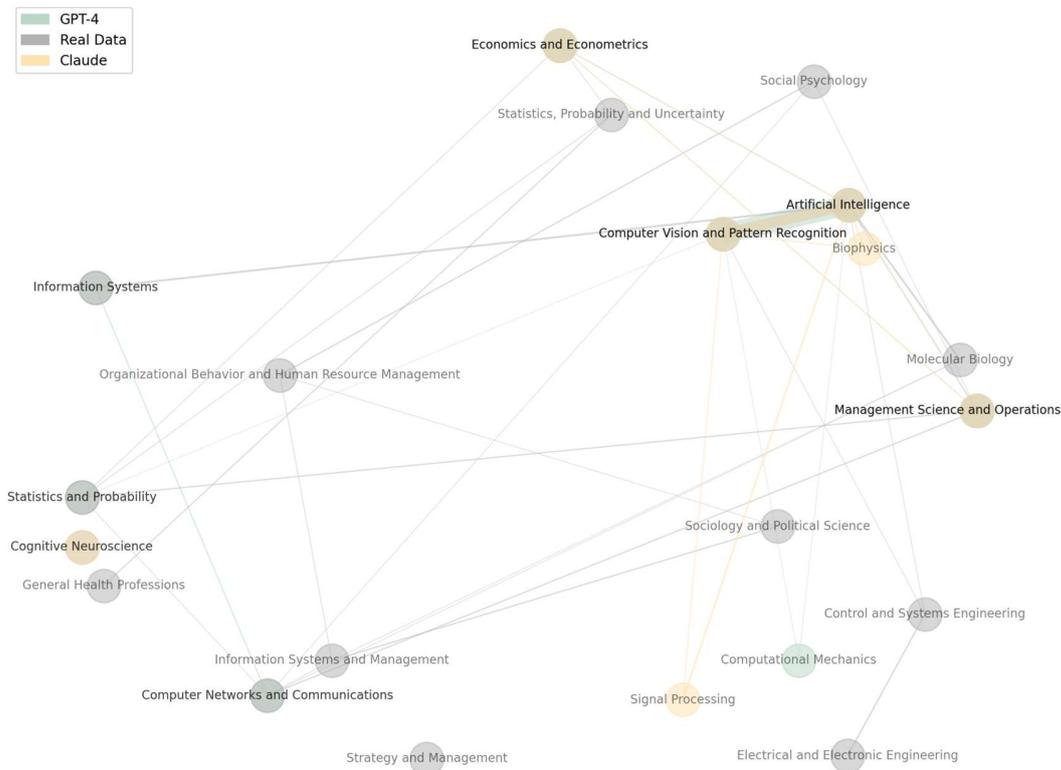

**Figure 3. The topic map of the domain of machine learning.**

In the field of machine learning, core topics such as Artificial Intelligence, Information Systems, and Pattern Recognition were consistently recognized and recommended by large language models. However, in practice, important machine learning literature also extends to applications in diverse fields, including organizational behavior, sociology, and political science. While ChatGPT demonstrates a tendency to focus on traditional computer science topics, Claude exhibits a broader perspective, more effectively capturing contributions from interdisciplinary subfields such as economics and cognitive neuroscience.

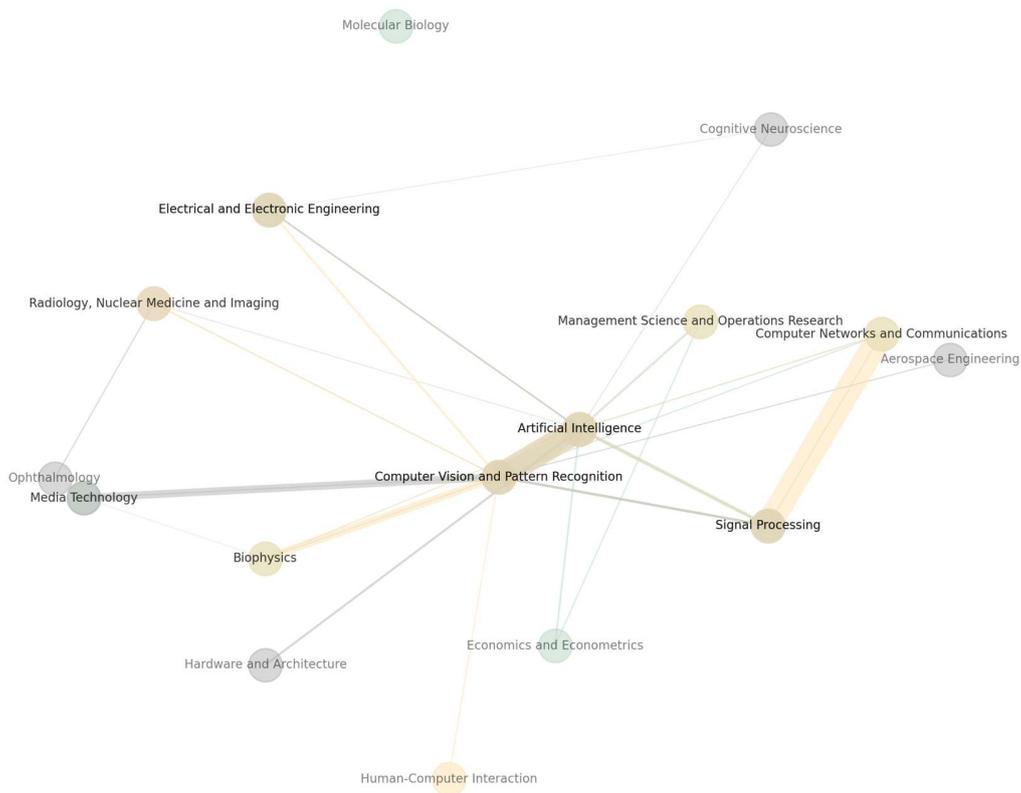

**Figure 4. The topic map of the domain of deep learning.**

In the field of deep learning, the recommended results align more closely with the real-world distribution. This may be due to the fact that most important deep learning literature is concentrated in the area of computer vision, resulting in less variation in the recommended papers. Both ChatGPT and Claude showed awareness of non-computer science fields, incorporating topics such as management science and biology. Notably, ChatGPT accurately identified important literature in the field of Media Technology, while Claude included some less-cited research from the field of human-computer interaction, deeming it worthy of consideration as important literature.

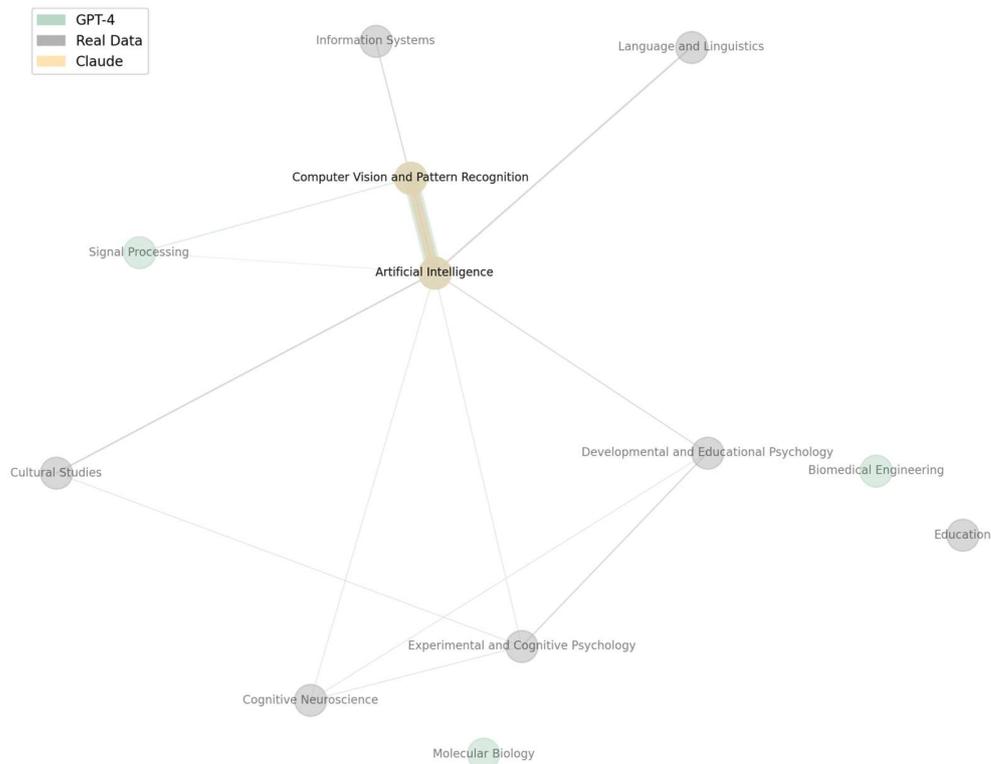

**Figure 5. The topic map of the natural language processing.**

The results in the field of natural language processing (NLP) closely resemble those in deep learning, with a strong emphasis on topics related to Pattern Recognition. Notably, both LLMs struggled to effectively recommend cross-disciplinary NLP papers, such as important studies in computational linguistics, educational psychology, and cultural studies. ChatGPT's attention to non-computer science fields remained limited to STEM disciplines, while Claude did not extend its focus to areas outside of computer science at all.

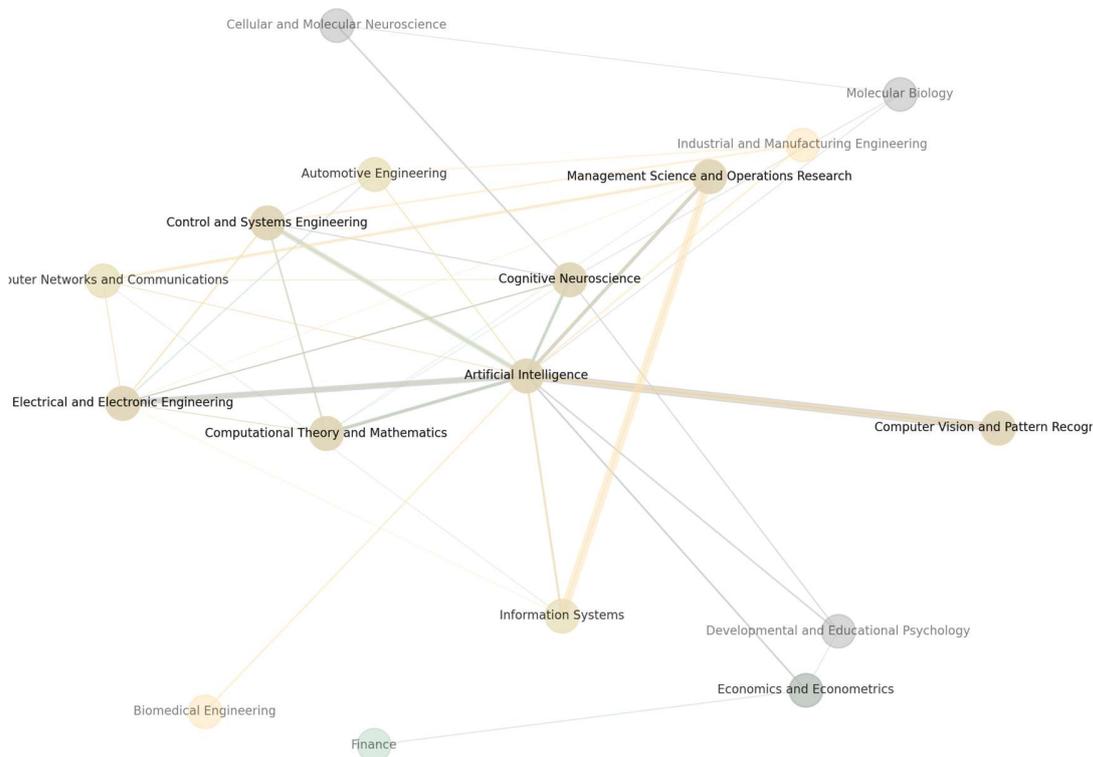

**Figure 6. The topic map of the domain of reinforcement learning.**

In the field of reinforcement learning, the range of topics covered by LLMs expanded significantly. Both models correctly recommended important papers in areas such as artificial intelligence, electrical engineering, and computational theory. Claude's focus was broader, with particular attention given to fields within engineering, including electrical engineering, automation control, and industrial production. In contrast, ChatGPT4's focus was more concentrated, primarily on artificial intelligence and computational theory, with some interest in econometrics.

By analyzing the differences between the topic distribution of LLM-recommended results and the real-world research topic distribution, we found that LLMs struggle to capture the interdisciplinary nature of research in artificial intelligence. They tend to focus more on topics within the computing domain. While LLMs can reliably provide a collection of important technical literature, they face challenges in recognizing specific applied fields.

Not only are they constrained by inherent limitations, which prevent them from identifying groundbreaking interdisciplinary applications as humans do, but their recommendations may also reflect internal biases based on their parameter settings, resulting in a preference for certain topics.

*Scholar Level*

All models' gender distributions are almost identical to the real situation, reflecting the dominance of men over women.

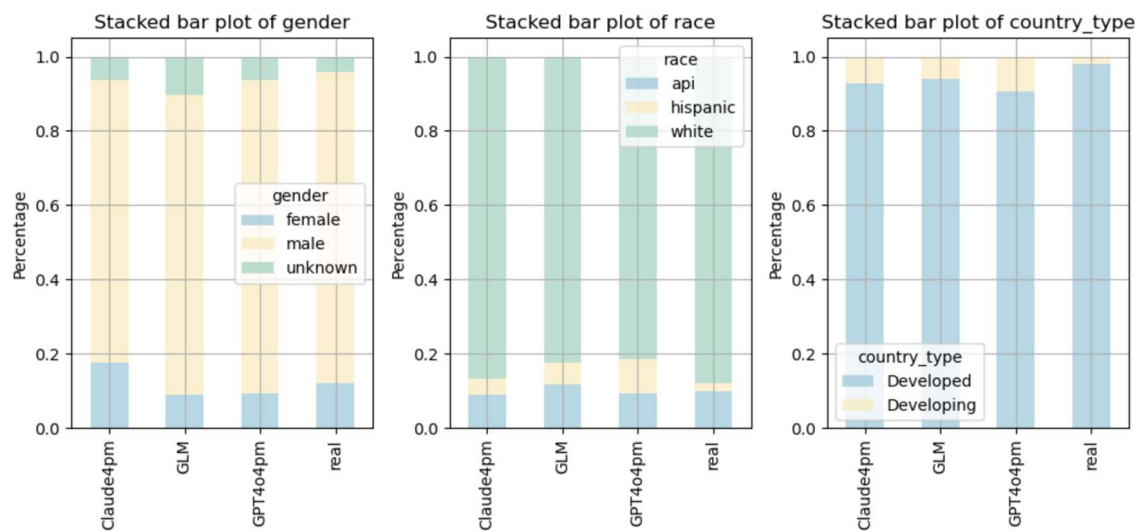

**Figure 7. Stacked bar chart of gender, race, and nationality distribution of authors recommended by different LLMs.**

Chi-square tests showed no significant difference from the real data, thus we reject hypotheses 2-4. However, the distribution of results from the three LLMs differed slightly from the real situation. The proportion of white authors in the recommendations was slightly lower than in reality for both GLM and ChatGPT4, while the proportions of Asian and Latino authors were slightly higher, particularly in ChatGPT's results.

Although not statistically significant, the proportion of authors from developing countries in the results returned by all three LLMs was higher than in the real data. These findings may suggest that the LLMs have undergone training designed to mitigate biases, allowing them to more comprehensively represent the global academic community.

**Table 3. The results of the null hypothesis tests.**

WHO GETS RECOMMENDED?

|   | Hypothesis | Results |
|---|---|---|
| 1 | LLM tends to recommend papers with higher citation counts. | Rejected |
| 2 | LLM tends to recommend male scholars. | Rejected |
| 3 | LLM tends to recommend white scholars. | Rejected |
| 4 | LLM tends to recommend scholars from developed countries. | Rejected |

*Robustness Checks*

The necessity of robustness testing arises from the need to ensure that a model's predictions remain stable and consistent under various conditions. Different packages and algorithms may exhibit inconsistencies or biases when handling complex data, so robustness testing is crucial to verify the reliability and credibility of the model's outputs. This is especially important in tasks such as race and gender prediction, where the data involved is sensitive. Ensuring that the model's results are not only valid for a specific dataset but also stable across broader datasets is essential. By comparing different models, robustness testing can uncover potential biases, data issues, or algorithmic limitations, offering valuable insights for future model optimization.

In this study, robustness testing for race and gender predictions was conducted to ensure that the final model's results remained consistent and reliable across different algorithms, thus minimizing misleading outcomes caused by the instability of any single approach. Given the potential instability in race and gender predictions using various Python packages, robustness testing was necessary to assess the reliability of the results. To perform the robustness check, two additional methods were employed for both race and gender prediction tasks. For race prediction, the Surgeo (*https://github.com/theonaunheim/surgeo*) and Ethnicolr (*https://github.com/appeler/ethnicolr*) packages were used. For gender prediction, the sexmachine package (*https://github.com/ferhatelmas/sexmachine*) was applied, alongside a combined approach using nameparser ([32]*https://github.com/derek73/python-nameparser*), nltk[36], and gender-guesser (*https://gender-guesser.com*). In this combined approach, nameparser was used for name extraction, nltk handled part of the gender predictions using its name dataset, and gender-guesser was employed for cases where nltk could not provide predictions. It is important to note that the sexmachine package is no longer maintained and does not support Python 3, so modifications were made to adapt it for this environment.

Robustness testing was carried out on data from the author recommendation task. After excluding cases where predictions were not possible due to data format issues or ambiguous names (e.g., "Andy," which could refer to either gender), the average Kappa values for each model were calculated and compared to the actual results. The table below (omitted here) presents the details of these comparisons.

The results indicated that gender predictions were notably stable, with full agreement across all models after excluding cases where predictions could not be made. In contrast, race prediction showed more variability. Assuming that a Kappa value greater than 0.6 indicates successful robustness, the results from the Surgeo package mostly passed, whereas the pred_fl_reg_name method performed poorly. This suggests that more stable methods for race prediction should be explored in future experiments to improve the robustness and accuracy of these predictions.

**Table 4. Performance of different LLMs with various settings and contexts.**

|  | Claude 3.5 | GLM 4 | GPT4o | Reality |
|---|---|---|---|---|
| Race (surgeo) | 0.66 | 0.56 | 0.58 | 0.68 |
| Race (pred_fl_reg_name) | 0.35 | 0.38 | 0.33 | 0.36 |
| Gender (sexmachine) | 1 | 1 | 1 | 1 |
| Gender (nltk & gender-gusser) | 1 | 1 | 1 | 1 |

## DISCUSSION

While large language models (LLMs) can provide usable recommendations after manual screening, their accuracy in specific literature recommendation tasks remains moderate. Their preferences tend to favor timeliness, collaboration, and incremental developments in research over disruptive innovations, with no other notable biases observed compared to real-world distributions. These findings highlight the need for continued refinement of LLMs and emphasize the ethical considerations in their application within academic contexts.

Our study distinguishes itself from existing research on LLMs in several important ways. While much of the current AI research in scientometrics focuses on tasks like abstract generation[37,38] and Thelwall et al. have conducted a general, large scale evaluation of LLM bias on academic paper database[39], we direct our attention to the underexplored area of literature recommendation. This shift not only broadens the scope of inquiry but also addresses practical scenarios that are increasingly relevant to scientific workflows. Unlike previous study on literature recommendation[40], which detect the difference between recommendation and 'reality' by generating reference lists for lately published articles, our approach prioritizes the comprehensive exploration of domain knowledge. By framing the task from the perspective of entry-level scholars or enthusiasts for AI fields, we align our research with real-world usage patterns. This makes our findings particularly valuable in understanding and predicting how biases in literature recommendation might influence the dissemination of existing disparity in science.

Moreover, our work takes a distinct approach to examining LLM biases. Instead of focusing broadly on human or societal dimensions, we narrow our analysis to citation recommendation within a specific domain. By employing quantitative methods rather than traditional experimental approaches in sociology or psychology[41], we provide a robust and data-driven perspective on the biases embedded in LLM-driven recommendations, contributing both to methodological innovation and practical relevance in this field.

Our study has several limitations that should be noted. First, the benchmark used in this research was manually curated by experts. While this ensures the selection of relevant and high-quality literature, it inherently limits the explanatory power and broader applicability of the findings. Second, the sample size of 50 may appear relatively small; however, this was determined with the potential issue of LLMs fabricating citations in mind. A larger sample size could result in an excessive number of fabricated references, complicating the analysis and potentially compromising the reliability of the results. Consequently, increasing the sample size further presents significant challenges. Third, the range of LLMs included in our study could be expanded to incorporate more emerging model architectures. This would provide a more comprehensive understanding of the biases and behaviors across a broader spectrum of models. Finally, our analysis is limited to the domain of computer science and artificial intelligence. This focus restricts the generalizability of our conclusions, as the observed patterns may not necessarily extend to other academic disciplines with different publication and citation dynamics. Future research could explore these aspects further, such as optimizing prompt design, expanding comparative indicators, and developing more varied experimental approaches. This would allow for a more comprehensive evaluation of the inequality impacts of LLMs, enhancing the understanding of their role in shaping academic landscapes.

## CONCLUSIONS

In this study, we examined the performance and potential biases of several large language models (LLMs) in literature recommendation tasks. We evaluated models such as Claude, ChatGPT, and GLM by assessing their ability to recommend key literature and scholars in fields like Artificial Intelligence and Machine Learning. Our findings reveal that the overall error rate of LLM recommendations varies, with Claude showing the best performance and GLM performing the worst. While LLMs generally recommended documents with lower citation counts compared to 'real' important literature, they still prefer highly-cited works to lower cited ones, suggesting a partial consideration of citation impact. The team sizes of the recommended documents aligned with real-world data, with most recommendations coming from small teams. Notably, LLMs displayed a significant preference for more recently published documents, particularly those from the past 20 years. The models tended to favor more conservative, developmental papers rather than highly disruptive ones. The proportions of different genders, races, and country types in the LLM-recommended results show no significant differences compared to real-world conditions, indicating that LLMs do not exhibit a preference for male, white, or scholars from developed countries, which contradicts known human discrimination patterns. Additionally, there was a noticeable "compensation" effect, where the models recommended a higher proportion of scholars from developing countries compared to the actual distribution. Although this difference was not statistically significant, it suggests that the large models may have been trained to mitigate biases, helping to more comprehensively represent the academic community. This could, to some extent, address issues of underrepresentation related to gender, race, and nationality in academia. By identifying and addressing potential biases in this task space, we envision that future LLM literature recommenders can provide relevant and fairer recommendation results and facilitate healthy and equitable exposure of research.

## ACKNOWLEDGMENTS

Yi Bu acknowledges the financial support from the National Natural Science Foundation of China (#72474009, #72104007, and #72174016).

# REFERENCES

1. Imran, M. & Almusharraf, N. Analyzing the role of ChatGPT as a writing assistant at higher education level: A systematic review of the literature. *CONT ED TECHNOLOGY* **15**, ep464 (2023).

2. Meyer, J. G. *et al.* ChatGPT and large language models in academia: opportunities and challenges. *BioData Mining* **16**, 20 (2023).

3. Thelwall, M. ChatGPT for complex text evaluation tasks. *Journal of the Association for Information Science and Technology* (2024),.

4. Ali, N. F., Mohtasim, M. M., Mosharrof, S. & Krishna, T. G. Automated Literature Review Using NLP Techniques and LLM-Based Retrieval-Augmented Generation. Preprint at https://doi.org/10.48550/arXiv.2411.18583 (2024).

5. Tan, Z. *et al.* Large Language Models for Data Annotation and Synthesis: A Survey. Preprint at https://doi.org/10.48550/arXiv.2402.13446 (2024).

6. Jin, H., Zhang, Y., Meng, D., Wang, J. & Tan, J. A Comprehensive Survey on Process-Oriented Automatic Text Summarization with Exploration of LLM-Based Methods. Preprint at https://doi.org/10.48550/arXiv.2403.02901 (2024).

7. Liang, W. *et al.* Can Large Language Models Provide Useful Feedback on Research Papers? A Large-Scale Empirical Analysis. *NEJM AI* **1**, AIoa2400196 (2024).

8. Zou, J. & Schiebinger, L. AI can be sexist and racist — it's time to make it fair. *Nature* **559**, 324–326 (2018).

9. Guleria, A., Krishan, K., Sharma, V. & Kanchan, T. ChatGPT: ethical concerns and challenges in academics and research. *The Journal of Infection in Developing Countries*

**17**, 1292–1299 (2023).

10. Garg, N., Schiebinger, L., Jurafsky, D. & Zou, J. Word embeddings quantify 100 years of gender and ethnic stereotypes. *Proceedings of the National Academy of Sciences* **115**, E3635–E3644 (2018).

11. Stokel-Walker, C. ChatGPT listed as author on research papers: many scientists disapprove. *Nature* **613**, 620–621 (2023).

12. Petiska, E. ChatGPT cites the most-cited articles and journals, relying solely on Google Scholar's citation counts. As a result, AI may amplify the Matthew Effect in environmental science. Preprint at https://doi.org/10.48550/arXiv.2304.06794 (2023).

13. Gallegos, I. O. *et al.* Bias and Fairness in Large Language Models: A Survey. *Computational Linguistics* **50**, 1097–1179 (2024).

14. Merton, R. K. The Matthew Effect in Science. *Science* **159**, 56–63 (1968).

15. Larivière, V. & Gingras, Y. The impact factor's Matthew Effect: A natural experiment in bibliometrics. *Journal of the American Society for Information Science and Technology* **61**, 424–427 (2010).

16. Larivière, V., Ni, C., Gingras, Y., Cronin, B. & Sugimoto, C. R. Bibliometrics: Global gender disparities in science. *Nature* **504**, 211–213 (2013).

17. Teich, E. G. *et al.* Citation inequity and gendered citation practices in contemporary physics. *Nat. Phys.* **18**, 1161–1170 (2022).

18. Lerman, K., Yu, Y., Morstatter, F. & Pujara, J. Gendered citation patterns among the scientific elite. *Proc. Natl. Acad. Sci. U.S.A.* **119**, e2206070119 (2022).

19. Hopkins, A. L., Jawitz, J. W., McCarty, C., Goldman, A. & Basu, N. B. Disparities

in publication patterns by gender, race and ethnicity based on a survey of a random sample of authors. *Scientometrics* **96**, 515–534 (2013).

20. Gomez, C. J., Herman, A. C. & Parigi, P. Leading countries in global science increasingly receive more citations than other countries doing similar research. *Nat Hum Behav* **6**, 919–929 (2022).

21. Tian, Y. & Bu, Y. Developed Countries Dominate Leading Roles in International Scientific Collaborations: Evidence from Scholars' Self-Reported Contribution in Publications. *Proceedings of the Association for Information Science and Technology* **61**, 1104–1106 (2024).

22. Lin, Z., Yin, Y., Liu, L. & Wang, D. SciSciNet: A large-scale open data lake for the science of science research. *Sci Data* **10**, 315 (2023).

23. Priem, J., Piwowar, H. & Orr, R. OpenAlex: A fully-open index of scholarly works, authors, venues, institutions, and concepts. Preprint at https://doi.org/10.48550/arXiv.2205.01833 (2022).

24. Zeng, A. *et al.* GLM-130B: An Open Bilingual Pre-trained Model. Preprint at https://doi.org/10.48550/arXiv.2210.02414 (2023).

25. Amin, M. M., Mao, R., Cambria, E. & Schuller, B. W. A Wide Evaluation of ChatGPT on Affective Computing Tasks. *IEEE Transactions on Affective Computing* **15**, 2204–2212 (2024).

26. Peng, K. *et al.* Towards Making the Most of ChatGPT for Machine Translation. Preprint at https://doi.org/10.48550/arXiv.2303.13780 (2023).

27. Brown, T. B. *et al.* Language Models are Few-Shot Learners. Preprint at https://doi.org/10.48550/arXiv.2005.14165 (2020).

28. Chen, L., Zaharia, M. & Zou, J. How is ChatGPT's behavior changing over time? Preprint at https://doi.org/10.48550/arXiv.2307.09009 (2023).

29. Sinha, A. *et al.* An Overview of Microsoft Academic Service (MAS) and Applications. in *Proceedings of the 24th International Conference on World Wide Web* 243–246 (Association for Computing Machinery, New York, NY, USA, 2015). doi:10.1145/2740908.2742839.

30. Beltagy, I., Lo, K. & Cohan, A. SciBERT: A Pretrained Language Model for Scientific Text. Preprint at https://doi.org/10.48550/arXiv.1903.10676 (2019).

31. Dong, Y., Ma, H., Shen, Z. & Wang, K. A Century of Science: Globalization of Scientific Collaborations, Citations, and Innovations. in *Proceedings of the 23rd ACM SIGKDD International Conference on Knowledge Discovery and Data Mining* 1437–1446 (ACM, Halifax NS Canada, 2017). doi:10.1145/3097983.3098016.

32. Kinney, R. *et al.* The Semantic Scholar Open Data Platform. Preprint at https://doi.org/10.48550/arXiv.2301.10140 (2023).

33. Bu, Y., Li, M., Gu, W. & Huang, W. Topic diversity: A discipline scheme-free diversity measurement for journals. *Journal of the Association for Information Science and Technology* **72**, 523–539 (2021).

34. Leydesdorff, L., Wagner, C. S. & Bornmann, L. Interdisciplinarity as diversity in citation patterns among journals: Rao-Stirling diversity, relative variety, and the Gini coefficient. *Journal of Informetrics* **13**, 255–269 (2019).

35. Wu, L., Wang, D. & Evans, J. A. Large teams develop and small teams disrupt science and technology. *Nature* **566**, 378–382 (2019).

36. Bird, S., Klein, E. & Loper, E. *Natural Language Processing with Python:*

*Analyzing Text with the Natural Language Toolkit*. (O'Reilly Media, Inc., 2009).

37. Walters, W. H. & Wilder, E. I. Fabrication and errors in the bibliographic citations generated by ChatGPT. *Sci Rep* **13**, 14045 (2023).

38. Qureshi, R. *et al.* Are ChatGPT and large language models "the answer" to bringing us closer to systematic review automation? *Syst Rev* **12**, 72 (2023).

39. Thelwall, M. & Kurt, Z. Research evaluation with ChatGPT: Is it age, country, length, or field biased? Preprint at https://doi.org/10.48550/arXiv.2411.09768 (2024).

40. Algaba, A. *et al.* Large Language Models Reflect Human Citation Patterns with a Heightened Citation Bias. Preprint at https://doi.org/10.48550/arXiv.2405.15739 (2024).

41. Manerba, M. M., Stańczak, K., Guidotti, R. & Augenstein, I. Social Bias Probing: Fairness Benchmarking for Language Models. Preprint at https://doi.org/10.48550/arXiv.2311.09090 (2024).

# APPENDIX

*'Real' Important Papers List*

| ML | DL | RL | NLP |
| --- | --- | --- | --- |
| Deep Residual Learning for Image Recognition | Deep Residual Learning for Image Recognition | Reinforcement Learning: An Introduction | Attention Is All You Need |
| Adam: A Method for Stochastic Optimization | Long Short-Term Memory | Human-level control through deep reinforcement learning | Glove: Global Vectors for Word Representation |
| U-Net: Convolutional | Deep learning | Deep learning in neural networks: An overview | A tutorial on hidden Markov |
| Networks for Biomedical Image Segmentation | | | models and selected applications in speech recognition |
| Gradient-based learning applied to document recognition | Very Deep Convolutional Networks for Large-Scale Image Recognition | Mastering the game of Go with deep neural networks and tree search | Learning Phrase Representations using RNN Encoder–Decoder for Statistical Machine Translation |
| ImageNet: A large-scale hierarchical image database | Gradient-based learning applied to document recognition | Reinforcement Learning: A Survey | BLEU: A Method for Automatic Evaluation of Machine Translation |
| ImageNet Classification with Deep Convolutional Neural Networks | Going deeper with convolutions | Advances in Neural Information Processing Systems 14 | WordNet: a lexical database for English |
| Optimization by Simulated Annealing | ImageNet classification with deep convolutional neural networks | Mastering the game of Go without human knowledge | Attention is All you Need |
| Genetic algorithms in search, optimization, and machine learning | Human-level control through deep reinforcement learning | Q-learning | ggplot2 |
| Scikit-learn: Machine Learning in Python | Faster R-CNN: Towards real-time object detection with region proposal networks | Continuous control with deep reinforcement learning | A translation approach to portable ontology specifications |
| Going deeper with convolutions | Learning Multiple Layers of Features from Tiny Images | Artificial intelligence: a modern approach | Pyramid Scene Parsing Network |
| Coefficient alpha and the internal structure of tests | Deep learning in neural networks: An overview | Reinforcement Learning: A Survey | Efficient Estimation of Word |
| | | | Representations in Vector Space |
| Meta-analysis in clinical trials | SegNet: A Deep Convolutional Encoder-Decoder Architecture for Image Segmentation | Simple statistical gradient-following algorithms for connectionist reinforcement learning | Back-Translation for Cross-Cultural Research |
| Structural equation modeling in practice: A review and recommended two-step approach. | Delving Deep into Rectifiers: Surpassing Human-Level Performance on ImageNet Classification | On-line Q-learning using connectionist systems | Sequence to Sequence Learning with Neural Networks |
| Regression Shrinkage and Selection Via the Lasso | Very Deep Convolutional Networks for Large-Scale Image Recognition | Predictive Reward Signal of Dopamine Neurons | Convolutional Neural Networks for Sentence Classification |
| ImageNet Large Scale Visual Recognition Challenge | Xception: Deep Learning with Depthwise Separable Convolutions | Asynchronous Methods for Deep Reinforcement Learning | RoBERTa: A Robustly Optimized BERT Pretraining Approach |
| Firm Resources and Sustained Competitive Advantage | Representation Learning: A Review and New Perspectives | Model-Agnostic Meta-Learning for Fast Adaptation of Deep Networks | Neural Machine Translation by Jointly Learning to Align and Translate |
| Support-vector networks | Sequence to Sequence Learning with Neural Networks | Playing Atari with Deep Reinforcement Learning | The NIST definition of cloud computing |
| Attention Is All You Need | A survey on deep learning in medical image analysis | Overcoming catastrophic forgetting in neural networks | Foundations of Statistical Natural Language Processing |
| LIBSVM: a library for support vector machines | Caffe: Convolutional Architecture for | Deep Reinforcement Learning with Double Q-Learning | Enriching Word Vectors with |
| Asymptotic and resampling strategies for assessing and comparing indirect effects in multiple mediator models | Fast Feature Embedding MobileNets: Efficient Convolutional Neural Networks for Mobile Vision Applications | Introduction to Reinforcement Learning | Subword Information

Language Models are Few-Shot Learners |
| Classification of Surgical Complications | PyTorch: An Imperative Style, High-Performance Deep Learning Library | Policy Gradient Methods for Reinforcement Learning with Function Approximation | Deep Contextualized Word Representations |
| ImageNet classification with deep convolutional neural networks | Dermatologist-level classification of skin cancer with deep neural networks | The neural basis of human error processing: Reinforcement learning, dopamine, and the error-related negativity. | Finding Structure in Time |
| Reinforcement Learning: An Introduction | Photo-Realistic Single Image Super-Resolution Using a Generative Adversarial Network | Evolving Neural Networks through Augmenting Topologies | Distributed Representations of Words and Phrases and their Compositionality |
| Dropout: a simple way to prevent neural networks from overfitting | Spatial Pyramid Pooling in Deep Convolutional Networks for Visual Recognition | Temporal difference learning and TD-Gammon | The coding manual for qualitative researchers |
| Consolidated criteria for reporting qualitative research (COREQ): a 32-item checklist for interviews and focus groups | Understanding the difficulty of training deep feedforward neural networks | Neural Architecture Search with Reinforcement Learning | Neural Machine Translation by Jointly Learning to Align and Translate |
| Rich Feature Hierarchies for Accurate Object Detection and Semantic Segmentation | Rectified Linear Units Improve Restricted Boltzmann Machines | Between MDPs and semi-MDPs: A framework for temporal abstraction in reinforcement learning | Machine learning in automated text categorization |
| Human-level control through deep reinforcement learning | Learning Deep Features for Discriminative Localization | Deep Reinforcement Learning: A Brief Survey | An algorithm for suffix stripping |
| Judgment under Uncertainty: Heuristics and Biases | PointNet: Deep Learning on Point Sets for 3D Classification and Segmentation | Apprenticeship learning via inverse reinforcement learning | The self-organizing map |
| Highly accurate protein structure prediction with AlphaFold | Image Super-Resolution Using Deep Convolutional Networks | Representation Learning with Contrastive Predictive Coding | A spreading-activation theory of semantic processing. |
| Principles and Practice of Structural Equation Modeling | Fully Convolutional Networks for Semantic Segmentation | Playing Atari with Deep Reinforcement Learning | An Introduction to Functional Grammar |
| User Acceptance of Computer Technology: A Comparison of Two Theoretical Models | Learning long-term dependencies with gradient descent is difficult | A general reinforcement learning algorithm that masters chess, shogi, and Go through self-play | WordNet: An Electronic Lexical Database |
| MrBayes 3.2: Efficient Bayesian Phylogenetic Inference and Model Choice Across a Large Model Space | Speech recognition with deep recurrent neural networks | Reinforcement learning in robotics: A survey | Features of similarity. |
| Rethinking the Inception Architecture for Computer Vision | A survey on Image Data Augmentation for Deep Learning | Learning to predict by the methods of temporal differences | Distributed Representations of Words and Phrases and their Compositionality |
| A tutorial on hidden Markov models and selected applications in speech recognition | Learning Spatiotemporal Features with 3D Convolutional Networks | A guide to deep learning in healthcare | Fundamentals of speech recognition |
| SMOTE: Synthetic Minority Over-sampling Technique | Batch Normalization: Accelerating Deep Network Training by Reducing Internal Covariate Shift | Grandmaster level in StarCraft II using multi-agent reinforcement learning | Compilers: Principles, Techniques, and Tools |
| Modern Applied Statistics with S | Beyond a Gaussian Denoiser: Residual Learning of Deep CNN for Image Denoising | Soft Actor-Critic: Off-Policy Maximum Entropy Deep Reinforcement Learning with a Stochastic Actor | ROUGE: A Package for Automatic Evaluation of Summaries |
| Data Mining: Practical Machine Learning Tools and Techniques | Learning Deep Architectures for AI | Markov games as a framework for multi-agent reinforcement learning | Momentum Contrast for Unsupervised Visual Representation Learning |
| C4.5: Programs for Machine Learning | Unsupervised Representation Learning with Deep Convolutional Generative Adversarial Networks | Reinforcement Learning | Effective Approaches to Attention-based Neural Machine Translation |
| Search and clustering orders of magnitude faster than BLAST | Deep Learning Face Attributes in the Wild | Regularized Evolution for Image Classifier Architecture Search | The Stanford CoreNLP Natural Language Processing Toolkit |
| Learning Multiple Layers of Features from Tiny Images | Generative adversarial networks | Dissociable Roles of Ventral and Dorsal Striatum in Instrumental Conditioning | Building a Large Annotated Corpus of English: The Penn Treebank |
| Educational Research: Planning, Conducting, and Evaluating Quantitative and Qualitative Research | The Unreasonable Effectiveness of Deep Features as a Perceptual Metric | By Carrot or by Stick: Cognitive Reinforcement Learning in Parkinsonism | A Comprehensive Grammar of the English Language |
| A Survey on Transfer Learning | Development and Validation of a Deep Learning Algorithm for Detection of Diabetic Retinopathy in Retinal Fundus Photographs | Prioritized Experience Replay | Opinion Mining and Sentiment Analysis |
| The value of the world's ecosystem services and natural capital | Accurate Image Super-Resolution Using Very Deep Convolutional Networks | A Comprehensive Survey of Multiagent Reinforcement Learning | Thumbs up? Sentiment classification using machine learning techniques |
| Power System Stability and Control | Real-Time Single Image and Video Super-Resolution Using an Efficient Sub-Pixel Convolutional Neural Network | SeqGAN: Sequence Generative Adversarial Nets with Policy Gradient | Neural Machine Translation of Rare Words with Subword Units |
| IQ-TREE: A Fast and Effective Stochastic Algorithm for Estimating Maximum-Likelihood Phylogenies | Enhanced Deep Residual Networks for Single Image Super-Resolution | Self-Critical Sequence Training for Image Captioning | Individual differences in working memory and reading |
| A Theoretical Extension of the Technology Acceptance Model: Four Longitudinal Field Studies | Joint Face Detection and Alignment Using Multitask Cascaded Convolutional Networks | Does gamification work?--a literature review of empirical studies on gamification | Efficient Estimation of Word Representations in Vector Space |
| Statistical Learning Theory | Deep Convolutional Neural Networks for Computer-Aided Detection: CNN Architectures, Dataset | DARTS: Differentiable Architecture Search | Principles and Practice in Second Language Acquisition |
| An introduction to ROC analysis | Characteristics and Transfer Learning PyTorch: An Imperative Style, High-Performance Deep Learning Library | Progressive Neural Architecture Search | Aspects of the Theory of Syntax |
| Geneious Basic: An integrated and extendable desktop software platform for the organization and analysis of sequence data | Asynchronous Methods for Deep Reinforcement Learning | Maximum entropy inverse reinforcement learning | A solution to Plato's problem: The latent semantic analysis theory of acquisition, induction, and representation of knowledge. |
| Qualitative research and case study applications in education | Fashion-MNIST: a Novel Image Dataset for Benchmarking Machine Learning Algorithms | Context and Behavioral Processes in Extinction | Interpreting Qualitative Data Methods for Analysing Talk, Text and Interaction |
| Deep Residual Learning for Image Recognition | Deep Residual Learning for Image Recognition | Addressing function approximation error in actor-critic methods | Attention Is All You Need |

*'Real' Important Author List*

| ML | DL | RL | NLP |
|---|---|---|---|
| Geoffrey E. Hinton | Shaoqing Ren | Koray Kavukcuoglu | Ilya Sutskever |
| Kaiming He | Ilya Sutskever | David Silver | Christopher D. Manning |
| Yoshua Bengio | Yann Lecun | Demis Hassabis | Oriol Vinyals |
| Robert Tibshirani | Christian Szegedy | Jimmy Ba | Quoc V. Le |
| Karl Friston | Alex Krizhevsky | Pieter Abbeel | Richard Socher |

## WHO GETS RECOMMENDED?

| Ross Girshick | Koray Kavukcuoglu | Sergey Levine | Ruslan Salakhutdinov |
|---|---|---|---|
| Shaoqing Ren | Diederik P. Kingma | Timothy P. Lillicrap | Ali Farhadi |
| Xiangyu Zhang | Dumitru Erhan | Ioannis Antonoglou | Jiawei Han |
| Andrew Zisserman | Scott Reed | Daan Wierstra | James L. Mcclelland |
| Donald B. Rubin | Yangqing Jia | Stig Petersen | Alex Graves |
| Trevor Hastie | Jeff Donahue | Matthew Botvinick | Tomáš Mikolov |
| Ilya Sutskever | Ignacio Arganda-Carreras | Richard S. Sutton | Łukasz Kaiser |
| Alan Yuille | Ian Goodfellow | Martin Riedmiller | Jakob Uszkoreit |
| Yann Lecun | Timothy P. Lillicrap | Volodymyr Mnih | Noam Shazeer |
| Peter M. Bentler | Daan Wierstra | Alexander Pritzel | Kenton Lee |
| Jian Sun | Stig Petersen | Çağlar Gülçehre | Aaron Courville |
| Li Fei-Fei | Jonathan Long | Thore Graepel | Liang-Chieh Chen |
| Trevor Darrell | Andrew Rabinovich | Michael J. Frank | David M. Blei |
| Vladimir Vapnik | Soumith Chintala | Laurent Sifre | Ashish Vaswani |
| Leo Breiman | Hugo Larochelle | Arthur Guez | Luke Zettlemoyer |
| Michael I. Jordan | Abdelrahman Mohamed | Michael L. Littman | Jason Weston |
| Christopher D. Manning | Nitish Srivastava | Andrei A. Rusu | L. R. Rabiner |
| Christian Szegedy | Honglak Lee | Nathaniel D. Daw | Ming-Wei Chang |
| Anil K. Jain | Kyunghyun Cho | Marc G. Bellemare | Kristina Toutanova |
| Ai Koyanagi | Trevor Back | Joel Veness | Walter Kintsch |
| Jitendra Malik | Simon Kohl | Shane Legg | Niki Parmar |
| Arnold B. Bakker | Benoit Steiner | Georg Ostrovski | Daniel Ramage |

WHO GETS RECOMMENDED?

| | | | |
|---|---|---|---|
| M. E. J. Newman | Michael Figurnov | Julian Schrittwieser | Keith Rayner |
| Jerome H. Friedman | Nal Kalchbrenner | Andrew G. Barto | Llion Jones |
| Jürgen Schmidhuber | George E. Dahl | Razvan Pascanu | Angela D. Friederici |
| Luc Van Gool | Clemens Meyer | Simon Osindero | Illia Polosukhin |
| Piotr Dollár | Iasonas Kokkinos | Helen King | Aidan N. Gomez |
| Jinde Cao | Stanislav Nikolov | Joëlle Pineau | Jacob Devlin |
| Stephen Boyd | Alykhan Tejani | John Schulman | Ron J. Weiss |
| Ewout W. Steyerberg | Razvan Pascanu | Andy Barto | Patrick Haffner |
| Kalyanmoy Deb | Gregory Chanan | Aja Huang | Kyunghyun Cho |
| Eduardo Bernabé | Jon Shlens | Amir Sadik | Dzmitry Bahdanau |
| David A. Kenny | Song Han | Peter Stone | Andrej Karpathy |
| David L. Donoho | Francisco Massa | Nicolas Heess | Susan Dumais |
| Bradley Efron | Victor Lempitsky | Jan Peters | Jianfeng Gao |
| Alex Krizhevsky | Aja Huang | Marc Lanctot | Devi Parikh |
| Andrew F. Hayes | Geert Litjens | Mehdi Mirza | George A. Miller |
| Oriol Vinyals | Amir Sadik | Dario Amodei | Tat‐Seng Chua |
| Claes Fornell | David Warde-Farley | Mohammad Norouzi | Jerry A. Fodor |
| Quoc V. Le | Matthew D. Zeiler | Chelsea Finn | Li Deng |
| Koray Kavukcuoglu | Thomas Unterthiner | Yael Niv | Marta Kutas |
| David Silver | Mehdi Mirza | Raia Hadsell | Kevin Murphy |
| Felix Akpojene Ogbo | Xavier Glorot | Satinder Singh | Samy Bengio |
| Nan M. Laird | Wojciech Samek | Doina Precup | Zhiheng Huang |
| Robert F. Engle | Nicolas Papernot | Kenji Doya | Peter Prettenhofer |